\documentclass[conference]{IEEEtran}
\IEEEoverridecommandlockouts

\usepackage{cite}
\usepackage{amsmath,amssymb,amsfonts}
\usepackage{algorithmic}
\usepackage{graphicx}
\usepackage{textcomp}
\usepackage{xcolor}
\def\BibTeX{{\rm B\kern-.05em{\sc i\kern-.025em b}\kern-.08em
    T\kern-.1667em\lower.7ex\hbox{E}\kern-.125emX}}

\usepackage{pgfplots}

\pgfplotsset{compat=1.18}

\usepackage{url}

\newcommand{\graphScale}{0.8}
\newcommand{\graphWidthPercent}{0.58}
\newcommand{\graphHeightPercent}{0.42}

\begin{document}

\title{Investigating the Temporal Dynamics of Cyber Threat Intelligence} 

\author{\IEEEauthorblockN{Angel Kodituwakku, Clark Xu, \\ Daniel Rogers, and David K. Ahn}
\IEEEauthorblockA{Centripetal Networks \\
Reston, VA, USA \\
cis-research@centripetal.ai}
\and
\IEEEauthorblockN{Errin W. Fulp}
\IEEEauthorblockA{\textit{Department of Computer Science} \\
\textit{Wake Forest University} \\
Winston-Salem, NC, USA \\
fulp@wfu.edu}
}

\maketitle

\begin{abstract}


Indicators of Compromise (IoCs) play a crucial role in the rapid detection and mitigation of cyber threats. However, the existing body of literature lacks in-depth analytical studies on the temporal aspects of IoC publication, especially when considering up-to-date datasets related to Common Vulnerabilities and Exposures (CVEs). This paper addresses this gap by conducting an analysis of the timeliness and comprehensiveness of Cyber Threat Intelligence (CTI) pertaining to several recent CVEs. The insights derived from this study aim to enhance cybersecurity defense strategies, particularly when dealing with dynamic cyber threats that continually adapt their Tactics, Techniques, and Procedures (TTPs). Utilizing IoCs sourced from multiple providers, we scrutinize the IoC publication rate. Our analysis delves into how various factors, including the inherent nature of a threat, its evolutionary trajectory, and its observability over time, influence the publication rate of IoCs. Our preliminary findings emphasize the critical need for cyber defenders to maintain a constant state of vigilance in updating their IoCs for any given vulnerability. This vigilance is warranted because the publication rate of IoCs may exhibit fluctuations over time. We observe a recurring pattern akin to an epidemic model, with an initial phase following the public disclosure of a vulnerability characterized by sparse IoC publications, followed by a sudden surge, and subsequently, a protracted period with a slower rate of IoC publication.

\end{abstract}

\section{Introduction } 

Effective cybersecurity defense necessitates the maintenance of current threat actor and attack vector information in the form of Indicators of Compromise (IoCs). These IoCs encompass IP addresses, port numbers, file hashes, domains, URLs, and IDS signatures, and can be deployed on firewalls, IDS, other hardware devices, and software agents to bolster proactive defense. Cybersecurity analysts rely on accurate and up-to-date IoCs for threat hunting, while penetration testers use them to verify the security of defense systems and human elements against these attack vectors.

However, accumulating an effective set of timely and actionable IoCs poses a multifaceted challenge. When a threat is initially acknowledged, it is common for not all relevant IoCs to be attributed to the associated Common Vulnerabilities and Exposures (CVEs) or even discovered yet. Some may already appear in emerging threat feeds reported by automated systems and cybersecurity researchers. Zero-day vulnerabilities, in particular, exemplify this issue. Their exploitation is not widespread initially; instead, they are deployed with precision against high-value targets, leaving a minimal network and system footprint, thus hindering their visibility to cyber threat intelligence (CTI) providers and researchers. This challenge is partly due to the rarity and expense of zero-day vulnerabilities. This initial slow attribution and discovery phase presents challenges for cyber defenders in constructing defense mechanisms against such threats. Security breaches that exploit these attack vectors may occur without detection until intelligence feeds catch up to include the newly identified IoCs.

As the vulnerability becomes better understood and proof-of-concept scripts and payloads become more accessible, widespread exploitation may rapidly increase, leading to the discovery of new IoCs. Moreover, as the threat evolves, attackers often adjust their tactics, techniques, and procedures (TTPs) to evade detection. Maintaining an up-to-date defensive posture necessitates a continuous process of adding, updating and removing IoCs from intelligence feeds, as they can are discovered, change or become obsolete throughout the IoC's lifecycle. Additionally, it is essential to recognize that various CTI providers specialize in different domains of intelligence. This specialization can influence their coverage percentage as well as the timeliness and accuracy of their IoC publications.

To uphold a comprehensive defense posture, cyber defenders must adopt a strategy of continuously sourcing Indicators of Compromise (IoCs) from a wide array of providers. However, the task of gathering and maintaining IoCs can be daunting due to the escalating volume and complexity of cyber attacks. Fortunately, CTI is available from a variety of sources, offering defenders insights into current and emerging threats while providing actionable IoCs.

For instance, AlienVault is an Open Source Threat Intelligence (OSINT) platform that is publicly accessible at no cost and regularly updated by a robust community. On the other hand, certain CTI providers, like Sophos and IBM X-Force, offer intelligence for a fee, often distinguished by their specialization, recency, and level of detail \cite{Li19}. It is worth noting that CTI providers may not necessarily offer the same type of intelligence; some focus on emerging threats, while others specialize in specific aspects like malware analysis. Even when providers offer similar intelligence types, the degree of common information, known as intelligence overlap, is typically quite limited, with recent studies revealing an average overlap of no more than 4.5\% \cite{Li19}. Consequently, defenders may need to rely on CTI from multiple providers to ensure adequate IoC coverage for effective cybersecurity. Evaluating the quality of these feeds becomes crucial in leveraging them effectively as a defense strategy \cite{Griffioen20}.

While acquiring CTI from multiple providers can enhance coverage, the challenge then shifts to managing this wealth of information over time. CTI is generally developed through proactive hunting or reactive investigation \cite{rfc9424}. However, it is rare to possess complete intelligence about a particular threat when an attack is initially observed. This complexity is compounded by attackers who may adapt their Tactics, Techniques, and Procedures (TTPs) in response to available CTI, thereby extending the applicability of their attacks \cite{Griffioen20, Schlette21}. Consequently, outdated intelligence becomes a concern \cite{Li19}. For instance, a survey found that only 52\% of respondents were satisfied with the timeliness of intelligence, and 46\% expressed dissatisfaction with the lack of removal of expired IoCs \cite{Brown21}. Thus, defenders must not only manage multiple CTI sources for broader coverage but must do so continually. Regrettably, there has been limited research dedicated to understanding the temporal dynamics of CTI, such as when new IoCs for a specific threat tend to be published. This knowledge gap presents a challenge, hindering the development of proactive strategies for effective IoC management. A more comprehensive understanding of the timing aspects of IoC discovery and coverage could lead to more resilient and adaptive cybersecurity practices that align better with the ever-evolving threat landscape. In essence, the challenges surrounding IoCs underscore the imperative need for ongoing research and innovation in the cybersecurity field to maintain effective defense mechanisms in an ever-changing threat landscape.

This paper provides some initial insight to the temporal dynamics associated with a variety of open-source and commercial CTI providers and cyber attacks. The timeliness of IoCs is discussed in \cite{Griffioen20}; however, studies were limited to open-source CTI and few IoC examples. A focus of this paper is when IoCs associated with a specific threat tend to be published. This information can help defenders understand the vigilance required to manage threats. The notion of the IoC publication rate is also introduced, which represents the number of IoCs published during a period of time. To provide some insight into the publication of IoCs, the IoCs associated with six different vulnerabilities were obtained for a period of time. For each vulnerability, the attribution of IoCs started when the associated vulnerability was first publicly known. Statistics about when IoCs were published were then calculated. These initial results will show that in several cases the publication of IoCs was similar to an epidemic model~\cite{Wang10}, where the IoC publication rate was initially slow, spiked, then slowed again resembling the susceptible, infectious, and recovered phases. Knowledge of the rate of which IoCs are published over time can help defenders better prepare their defenses.

The remainder of this paper is structured as follows. Section 2 furnishes essential background information on CTI, including exploration of the different types of IoCs, cybersecurity provider specializations and categorization of feed sources. Furthermore, we discuss the widely accepted vulnerability identifiers, organizations that maintain information related to these vulnerabilities. In section 3 we discuss our IoC dataset and rate analysis. For each CVE analyzed we provide an IoC coverage graph and our observations. Finally, Section 4 encapsulates our findings and conclusions drawn from this research. We also outline potential avenues for future research that can extend and build upon the insights gleaned from this study.

\section{Background on Cyber Threat Intelligence}

CTI plays a critical role in cyber defense. They can be used to proactively block threats, detect intrusions, aid in threat hunting, and are crucial for incident response. Sharing IoCs fosters collective cyber defense. CTI sources can be broadly classified into two categories: open-source (free) and commercial (non-free). Open Source Threat Intelligence (OSINT) falls into the free category and typically comprises public lists of indicators, such as those from Abuse.ch, AlienVault, and DShield. In contrast, commercial CTI is available for a fee, often in the form of a subscription service. This type of intelligence is known for its high accuracy and specialization. It's important to note that CTI can also be accessible through sharing networks, such as U.S.~Cybersecurity and Infrastructure Security Agency (CISA) Automated Indicator Sharing (AIS) infrastructure \cite{AIS23}. This arrangement doesn't fall into the commercial or open-source categories, as intelligence is shared exclusively among group members based on shared interests for the sake of confidentiality.

CTI can be further categorized based on how the information is presented, which ranges from analytical trend reports in essay formats, suitable for security analysts, to machine-readable feeds, more appropriate for automated detection and blocking. Additionally, CTI providers may specialize in specific types of intelligence, like VirusShare.com, which publishes malware file hashes, or offer IoCs for various threat types under a multitude of categories, such as AlienVault.

Unfortunately, there is often little commonality between the CTI feeds, and as a result, defenders will need to consume CTI from as many sources as possible to provide the most comprehensive defense. Given the need to manage multiple CTI from different sources, understanding the quality is important and has been the subject of recent research \cite{Griffioen20,Schlette21}. Intelligence quality is often evaluated using multiple metrics, and two common measures are volume and timeliness \cite{Griffioen20}. Intelligence volume measures the total number of IoCs appearing over the measurement interval. Therefore, defenders would prefer high volume feeds as long as the IoCs reported are exclusive and free of false positives. CTI timeliness relates to the time between threat discovery and IoC publication \cite{Bouwman20}. IoC timeliness, also known as latency, is critical for effective defense posture. For example it has been reported that short latency is critical for successfully defending against phishing and spam \cite{Ramachandran07}. 



CTI can be associated with Common Vulnerability and Exposure (CVE) and National Vulnerability Database (NVD) reports. CVE is a list of publicly disclosed computer security flaws that is managed by the MITRE corporation with funding from the Cybersecurity and Infrastructure Security Agency (CISA), which is part of the U.S. Department of Homeland Security. Once a new vulnerability is discovered, it can be submitted to a CVE program participant for review. Once properly vetted, the vulnerability assigned a unique CVE number so the specific vulnerability can be easily identified in the future. The CVE and NVD information is then made public with the intent to improve defenses.

Typically, a CVE is generated when a vulnerability becomes publicly known. Subsequently, the NVD can augment the CVE entry with additional details, including severity assessments and mitigation advisories (noted as a CVE \textit{republish}). It is noteworthy and will be shown in the next section, that the majority of IoCs linked to a CVE surface after the CVE identifier has been assigned. However, it is important to recognize that IoCs specific to a particular CVE may still appear on intelligence feeds prior to this designation. This early appearance can be attributed to various factors, such as the misattribution of IoCs to a different CVE, attackers reusing available resources like botnets, suspicious domains and IP addresses reported by security providers' honeypots or through user reports on OSINT sources before the CVE designation, further contributing to this dynamic landscape of threat intelligence.

\section{IoC Rate Analysis }

We investigate the IoC rate of six CVEs (CVE-2023-34362, CVE-2023-35078, CVE-2023-37470, CVE-2023-21409, CVE-2023-2868, and CVE-2023-391431)\footnote{CVE details (including CVSS scores) available at \url{https://www.cve.org/}}. The Common Vulnerability Scoring System (CVSS) has designated these CVEs as ``critical'' due to their ease of attack and potential damage. IoCs for these CVEs encompass IP addresses and domain names sourced from 16 distinct Cyber Threat Intelligence (CTI) providers. This approach maximizes IoC coverage, as there is minimal overlap between these various sources.

It is worth noting that the names of the CTI providers cannot be disclosed due to contractual obligations. Each graph starts on ``Day 0,'' which corresponds to the CVE disclosure date, and continues until IoC coverage reaches 100\% for all known IoCs related to the respective CVE. This encompasses data from all sources considered during the analysis.

\subsection{IoC Coverage Observations }

\begin{figure}
\begin{center}
\begin{tikzpicture}[scale = \graphScale]
\begin{axis}[
    width=\graphWidthPercent\textwidth,
    height=\graphHeightPercent\textwidth,
    const plot,  
    no marks,  
    title=IoC Coverage for CVE-2023-34362 (MOVEit),
    xlabel=Days (since CVE published for this threat),  
    ylabel=Coverage percentage,  
]
\addplot[blue, very thick]  table[col sep=semicolon, x=index, y expr=\thisrowno{3}/149*100] {CVE-2023-34362_pcve.dat};
\end{axis}
\end{tikzpicture} 
\caption{IoC coverage (percentage of the 149 unique IoCs published) for CVE-2023-34362 over 75 successive days.}
\label{fig:cve-2023-34362}
\end{center}
\end{figure}
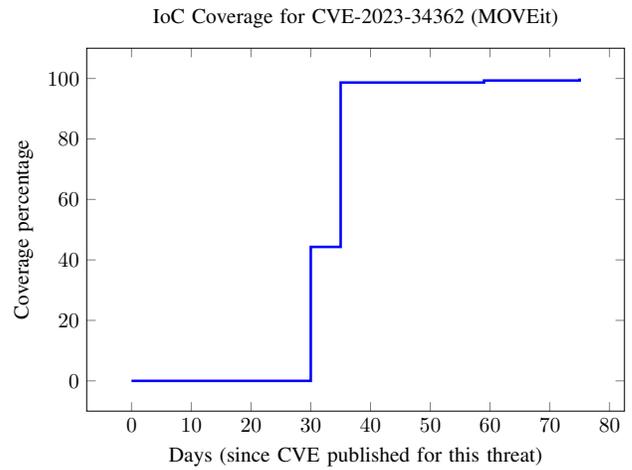

CVE-2023-34362 is a SQL injection vulnerability found in the MOVEit Transfer web application that allows an unauthenticated attacker to gain access to MOVEit Transfer's database. Depending on the database engine, an attacker may be able to infer information about the database and execute different SQL statements; as a result, the vulnerability was assigned a CVSS score of 9.8 (critical). The CVE and NVD were published on June 3, 2023 and last updated on June 23. Thirty days after the CVE, 149 different IPv4 address IoCs were published during the next 46 days. As seen in Figure~\ref{fig:cve-2023-34362}, the first set of IoCs published represented 44\% of IoCs that would be known by the end of the observation period. However 5 days after the initial IoC publication, the number of IoCs more than doubled and the coverage percentage increased to 98\%. This spike in the IoC publication rate was due to the discovery of a new set of IPv4 addresses associated with the vulnerability. An additional 24 days then passed until the remaining IoCs were published.

\begin{figure}
\begin{center}
\begin{tikzpicture}[scale = \graphScale]
\begin{axis}[
    width=\graphWidthPercent\textwidth,
    height=\graphHeightPercent\textwidth,
    const plot,  
    no marks,  
    title=IoC Coverage for CVE-2023-35078 (EPMM),
    xlabel=Days (since CVE published for this threat),  
    ylabel=Coverage percentage,  
]
\addplot[blue, very thick] table[col sep=semicolon, x=index, y expr=\thisrowno{3}/23*100] {cve-2023-35078_pcve.dat};
\end{axis}
\end{tikzpicture} 
\caption{IoC coverage (percentage of the 23 unique IoCs published) for CVE-2022-35078 over 49 successive days.}
\label{fig:cve-2023-35078}
\end{center}
\end{figure}
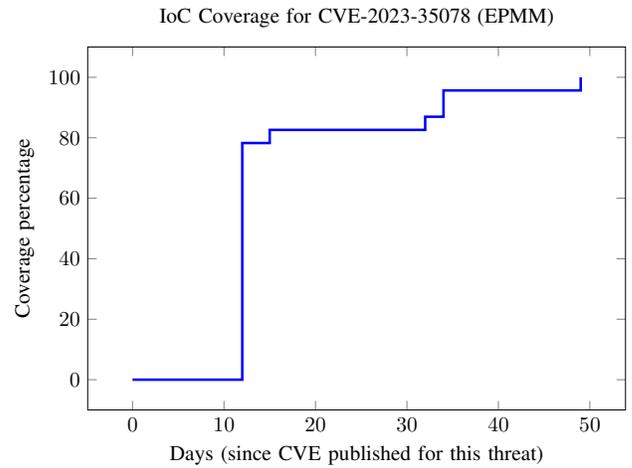

A similar IoC growth pattern was observed for CVE-2023-35078 and is depicted in Figure~\ref{fig:cve-2023-35078}. This CVE describes an authentication bypass with Ivanti EndPoint Manager Mobile (EPMM) that allows remote attackers to obtain personal information, add administrative accounts, and change the configuration. 
As a result of the possible damage, this vulnerability was assigned a CVSS score of 9.8 (critical).
The CVE and NVD were originally published on July 25, 2023 then last modified on August 4, 2023. Starting on August 6 (12 days after the CVE), 23 unique IoCs were associated with this vulnerability over a period of 38 days. 
Initially 78\% of the IoCs were available. The IoC publication rate was relatively small after this initial spike, averaging one new IoC every 12 days until coverage was complete. 

\begin{figure}
\begin{center}
\begin{tikzpicture}[scale = \graphScale]
\begin{axis}[
    width=\graphWidthPercent\textwidth,
    height=\graphHeightPercent\textwidth,
    const plot,  
    no marks,  
    title=IoC Coverage for CVE-2023-37470 (Metabase),
    xlabel=Days (since CVE published for this threat),  
    ylabel=Coverage percentage,  
]
\addplot[blue, very thick]  table[col sep=semicolon, x=index, y expr=\thisrowno{3}/63*100] {cve-2023-37470_pcve.dat};
\end{axis}
\end{tikzpicture} 
\caption{IoC coverage (percentage of the 63 unique IoCs published) for CVE-2023-37470 over 55 successive days.}
\label{fig:cve-2023-37470}
\end{center}
\end{figure}
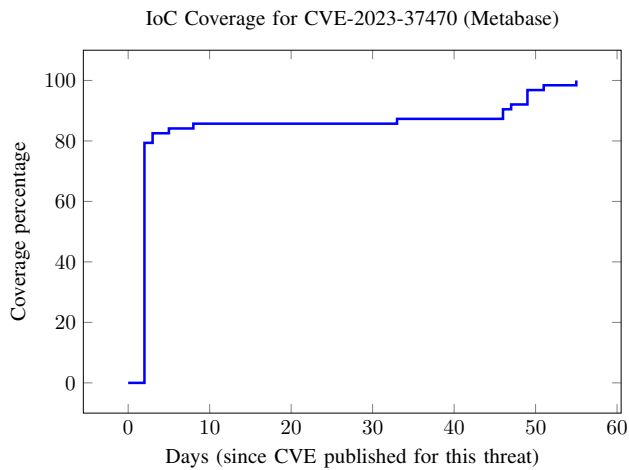

CVE-2023-37470 also experienced an large IoC coverage increase, as shown in Figure~\ref{fig:cve-2023-37470}. The vulnerability allows remote code execution on certain Metabase servers, which is an open-source business intelligence and analytics platform. It has been assigned a CVSS score of 9.8 (critical). CVE and NVD were originally published on August 4, 2023, the updated 5 days later. Starting two days after the CVE, 63 unique IoCs were published over the following 54 days. Over 79\% of the IoC were initially published and few updates occurred over the next 48 days. The remaining unique IoCs were the published average rate of one IoC per day over the last ten days until coverage was complete.

\begin{figure}
\begin{center}
\begin{tikzpicture}[scale = \graphScale]
\begin{axis}[
    width=\graphWidthPercent\textwidth,
    height=\graphHeightPercent\textwidth,
    const plot,  
    no marks,  
    title=IoC Coverage for CVE-2023-21409 (Axis),
    xlabel=Days (since CVE published for this threat),  
    ylabel=Coverage percentage,  
]
\addplot[blue, very thick]  table[col sep=semicolon, x=index, y expr=\thisrowno{3}/16*100] {cve-2023-21409_pcve.dat};
\end{axis}
\end{tikzpicture} 
\caption{IoC coverage (percentage of the 16 unique IoCs published) for CVE-2023-21409 over 25 successive days.}
\label{fig:cve-2023-21409}
\end{center}
\end{figure}
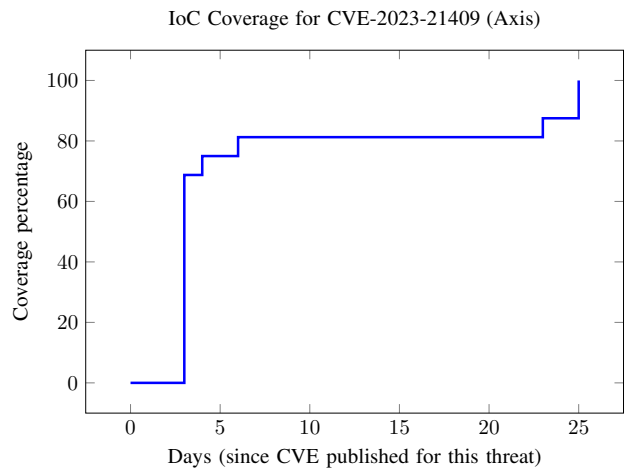

CVE-2023-21409 experienced an initial sudden increase in the IoCs published, as shown in Figure~\ref{fig:cve-2023-21409}. The vulnerability is associated with Axis License Plate Verifier application. Due to insufficient file permissions, attackers can gain access to unencrypted administrator credentials allowing the configuration of the application, and as a result, the vulnerability has a CVSS score of 9.8 (critical). CVE and NVD were originally published on August 3, 2023, then updated 5 days later. Three days after the CVE, 16 unique IoCs were published over 55 days. Over 68\% of the IoC were initially published and few updates occurred over the next 20 days. Afterwards the IoC publication rate changes to 1.5 IoCs per day until the remaining IoCs are published.

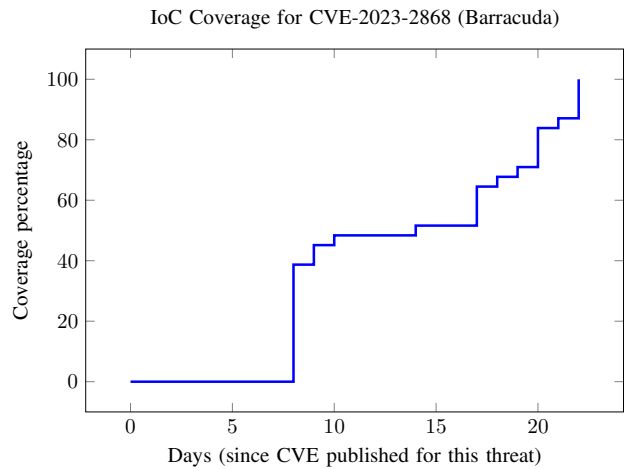
\begin{figure}
\begin{center}
\begin{tikzpicture}[scale = \graphScale]
\begin{axis}[
    width=\graphWidthPercent\textwidth,
    height=\graphHeightPercent\textwidth,
    const plot,  
    no marks,  
    title=IoC Coverage for CVE-2023-2868 (Barracuda),
    xlabel=Days (since CVE published for this threat),  
    ylabel=Coverage percentage,  
]
\addplot[blue, very thick]  table[col sep=semicolon, x=index, y expr=\thisrowno{3}/31*100] {cve-2023-2868_pcve.dat};
\end{axis}
\end{tikzpicture} 
\caption{IoC coverage (percentage of the 31 unique IoCs published) for CVE-2023-2868 over 24 successive days.}
\label{fig:cve-2023-2868}
\end{center}
\end{figure}

While the previous CVEs had a large initial number of published IoCs (over 50\%), the last two examples  considered (CVE-2023-2868 and CVE-2023-39143) had a smaller initial IoC publication rate.
CVE-2023-2868 is a remote command injection vulnerability that exists in the Barracuda Email Security Gateway. The vulnerability is due to not properly sanitizing tar file processing and resulted in a CVSS score of 9.8 (critical). The CVE and NVD were originally published on May 24, 2023, while the NVD was later updated on June 1, 2023. As seen in Figure \ref{fig:cve-2023-2868}, 38\% of the 31 IoCs were initially published on May 31, 2023, which is 8 days after the CVE. 
On average one new IoC was published every 3 days until the 17\textsuperscript{th} day, 3 new IoCs were published on average per day. This higher IoC publication rate continued until the 23\textsuperscript{rd} day, when all of the unique IoCs (known at that point in time) were published, resulting in the 100\% coverage of these IoCs.

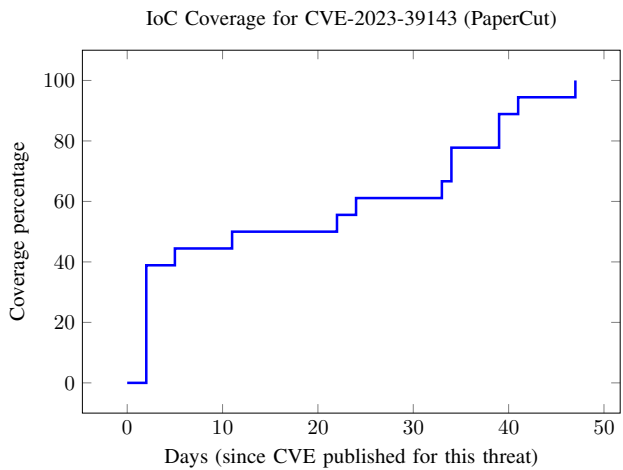
\begin{figure}
\begin{center}
\begin{tikzpicture}[scale = \graphScale]
\begin{axis}[
    width=\graphWidthPercent\textwidth,
    height=\graphHeightPercent\textwidth,
    const plot,  
    no marks,  
    title=IoC Coverage for CVE-2023-39143 (PaperCut),
    xlabel=Days (since CVE published for this threat),  
    ylabel=Coverage percentage,  
]
\addplot[blue, very thick]  table[col sep=semicolon, x=index, y expr=\thisrowno{3}/18*100] {cve-2023-39143_pcve.dat};
\end{axis}
\end{tikzpicture} 
\caption{IoC coverage (percentage of the 18 unique IoCs published) for CVE-2023-39143 over 47 successive days.}
\label{fig:cve-2023-39143}
\end{center}
\end{figure}

A similar publication pattern occurred for CVE-2023-39143, as depicted in Figure~\ref{fig:cve-2023-39143}. CVE-2023-39143 is a path traversal vulnerability associated with PaperCut NG and PaperCut MF software. The vulnerability allows an attackers to upload, read, or delete arbitrary files and can lead to remote code execution when external device integration is enabled. As a result this vulnerability has a CVSS score of 9.8 (critical). Both the CVE and NVD listings were originally published on August 4, 2023 and were subsequently updated August 8. Starting two days after the CVE, 7 unique IoCs were initially published, which is a relatively small initial spike. A more regular IoC publication rate of one new IoC every 4 days (on average) occurred until coverage was complete.

These CVE-based examples provide insight to how the number of IoCs published (IoC coverage) for a vulnerability changes over time. In most cases, there is a delay from the CVE publication until the first IoC is published. The initial phase is often followed by a spike in IoC publications, then the remaining IoCs are published at a slower rate during the last phase. This growth pattern is similar  to the three stages of the well-known Susceptible-Infected-Removed (SIR) model used to represent epidemic outbreaks and malware propagation~\cite{Wang10}. The early stages of IoC discovery is similar to the \textit{susceptible stage} of the SIR model, where the IoC publication rate remains low due to factors like limited exploit availability, lack of developed attack Tactics, Techniques, and Procedures (TTPs), and limited defender observation. The second phase is akin to the \textit{infectious stage} in of the epidemic model. The IoC rate surges due to automated exploits, commercialization, and targeted campaigns. The last phase, resembles the \textit{recovered stage} in the SIR model. In this case the IoC publication rate slows as defense techniques mature, awareness grows, and vulnerabilities are mitigated. Understanding these dynamics is crucial for informed cybersecurity decisions.
With this insight, defenders can be more aware of the attack stage and possibly deploy defenses more effectively.

\section{Conclusions and Future Work}

Third-party Cyber Threat Intelligence (CTI) can provide cyber defenders a scalable and effective means of acquiring actionable Indicators of Compromise (IoCs). Furthermore, given the number and scale of cyber attacks occurring at any given point in time, it is expected that defenders will acquire CTI (also referred to as a feed) from more than one CTI provider. Given this situation, where feed aggregates are likely to be used, it is helpful to understand when IoCs tend to be published. This provides defenders a notion of how vigilant they should be regarding maintaining up to date IoCs. 

This paper provided some initial insight on when IoCs were published for 6 different vulnerabilities. In several cases the publication rate of IoCs resembled an epidemic model that consists of a slow initial stage, a sudden spike, followed by a final slow stage. It is plausible that these publication stages coincide with the development, release, and defense of a vulnerability.

While these results are interesting, continued investigation of how IoC coverage develops is needed. More CVE examples should be analyzed to better understand the circumstances that cause the IoC publication rate to change, and in certain cases resemble an epidemic model. It might also be possible to infer attacker TTPs based on how the publication rate changes. In addition, considering the \textit{expiration} of IoCs (IoCs that are no longer observed) could give an understanding of IoC churn. Therefore, additional temporal insight about IoC coverage would help cyber defenders establish better defenses, both in terms of suitability and efficiency.

\bibliographystyle{IEEEtran}


\end{document}